\documentclass[journal]{IEEEtran}

\usepackage{amsmath,amsfonts,amssymb}
\usepackage{cite}
\usepackage[ruled,vlined,linesnumbered]{algorithm2e}
\usepackage{array}
\usepackage{booktabs}
\usepackage{multirow}
\usepackage{textcomp}
\usepackage{url}
\usepackage{graphicx}
\usepackage{xcolor}
\usepackage{tikz}
\usetikzlibrary{arrows.meta,positioning,fit,calc}
\SetKwInOut{Input}{Input}
\SetKwInOut{Output}{Output}

\usepackage{balance}

\begin{document}

\title{BACC: Budget-Aware Calibration and Control for Horizontal Autoscaling}

\author{Fan Liu, Behrooz Farkiani, Guanqi Li, and Patrick Crowley%
\thanks{The authors are with Washington University in St. Louis, St. Louis,
USA (e-mail: \{fan.liu, b.farkiani, guanqi, pcrowley\}@wustl.edu).}%
\thanks{This work was supported by NSF CNS Award 2213672.}}

\maketitle

\begin{abstract}
  A fixed-period service-level objective (SLO) defines a violation budget that an autoscaler can spend over the period, rather than a constraint to satisfy at every decision. Yet autoscalers commonly over-provision to avoid violations at every decision, which wastes resources. We present \textbf{BACC}, a budget-aware autoscaler that tracks the violation rate throughout the SLO period, and evaluate it via both simulation and system execution. BACC over-provisions when little budget remains and accepts some violations when much remains. It combines online conformal calibration of workload forecasts with observed SLO outcomes. BACC supports different forecasters and accommodates different SLO metrics through configurable replica sizing. We evaluate BACC on five serverless function traces under latency and CPU SLOs with 95\% and 99\% compliance targets. Via simulation, we show that BACC satisfies all 20 evaluated combinations, compared with 16 for the Horizontal Pod Autoscaler (HPA), the built-in autoscaler in Kubernetes. In those 16 combinations, BACC reduces mean provisioned replicas by 10.7--39.3\% for latency and 13.5--55.5\% for CPU. Experiments testing individual features show that fixed over-provisioning leaves 42--100\% of the violation budget unused, while budget pacing reduces replicas by up to 35\% and still meets the SLO. Replacing the primary forecaster with the recently proposed pretrained time-series model Chronos-2 preserves compliance in all 20 combinations. Beyond simulation, we also execute the same traces across multiple Kubernetes runs to both validate the simulation results and deepen some conclusions, particularly with bursty request patterns. On regular trace~B, both controllers comply, and BACC's mean paired replica reduction is 16.7\%. On bursty trace~D, BACC complies in 3/3 repetitions, whereas HPA complies in none. All the code used in the paper will be made publicly available upon publication.

\end{abstract}


\section{Introduction}
\label{sec:introduction}

Cloud services require computing resources, such as CPU cores and memory, to serve workloads that vary over time~\cite{rzadca2020autopilot}. Efficient resource management must balance the cost of these resources with meeting service-level objectives (SLOs). Provisioning for peak demand can protect against workload surges, but leaves excess capacity during periods of lower demand~\cite{shen2011cloudscale}. Autoscaling addresses this inefficiency by automatically adjusting resource capacity as workloads change. Vertical scaling changes the resources allocated to existing instances, while horizontal scaling adds or removes service instances~\cite{rzadca2020autopilot}. Platforms such as Kubernetes and Borg support autoscaling~\cite{kubernetes_hpa_docs,burns2016borg}, and the autoscaler determines when to change capacity and by how much.

We consider containerized services that scale by changing the number of service instances, or replicas. The SLO is evaluated over a fixed period with predefined start and end times, such as one month: the fraction of requests or measurement intervals that violate the SLO must remain at or below a specified limit. This allowed fraction constitutes a violation budget; for example, a 99\% compliance target permits at most 1\% violations over the period. The autoscaler's goal is to minimize the average number of provisioned replicas while keeping the violation rate over the full period within this limit.

Meeting this objective requires acting before demand arrives. After an autoscaler observes demand and requests additional replicas, those replicas must be scheduled, started, and made ready before serving requests. Reactive policies such as a typical Kubernetes Horizontal Pod Autoscaler (HPA) configuration can respond only after a demand increase becomes visible. Predictive autoscalers can act earlier by provisioning from a workload forecast~\cite{predictive_hpa_docs,urgaonkar2008agile,zou2023optscaler}, but forecasts can be wrong. The system also adds uncertainty, such as variable startup delay. To protect the SLO, autoscalers still often over-provision~\cite{rzadca2020autopilot,liu2024jiagu,dehigama2026spandana}. A 2025 industry report on more than 2,100 organizations found that average CPU utilization in Kubernetes clusters was only 10\%~\cite{castai2025benchmark}. Over-provisioning wastes resources, while under-provisioning causes SLO
violations~\cite{dehigama2026spandana}.

A fixed-period SLO gives the autoscaler a violation budget to spend, not a limit to satisfy at every decision. The right amount of over-provisioning therefore depends on how much budget remains. When most of the budget is spent, the autoscaler cannot afford more violations. It should over-provision to absorb forecast and system uncertainty. When much budget remains, it can accept some violations and save replicas. A fixed amount of over-provisioning cannot make this trade-off. We call the risk of under-provisioning accepted at each decision the \emph{risk tolerance}. The SLO requirement stays fixed. Only the risk tolerance of each decision changes with budget use.

\begin{figure*}[!t]
  \centering
  \includegraphics[width=0.96\textwidth]{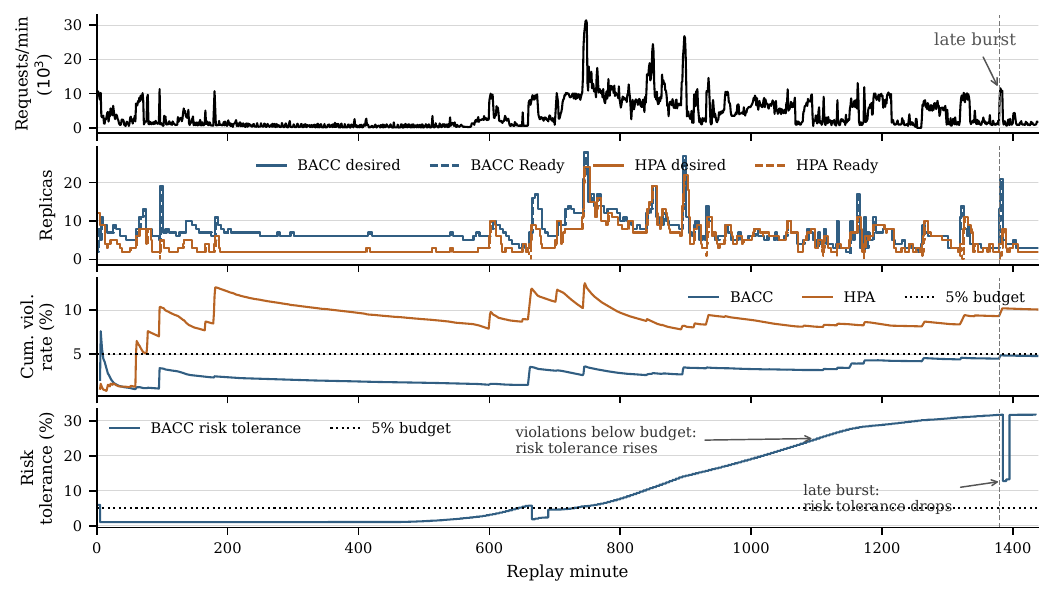}
  \caption{BACC and HPA on bursty trace~D (24-hour Kubernetes replay). BACC starts at its lowest risk tolerance because warm-up violations keep its cumulative violation rate above 5\% until about minute~440; the plotted rate excludes this 5-minute warm-up. Once violations fall below the budget, BACC raises its risk tolerance and spends the budget. When the late burst arrives near minute~1,380 (dashed line), the budget is nearly exhausted, so BACC lowers its risk tolerance and adds replicas. BACC ends within the budget, whereas HPA exceeds it early and remains above it. Desired replicas include pending scale-ups; dashed lines show Ready replicas.}
  \label{fig:k8s_trace_d_dynamics}
\end{figure*}

We present \textbf{BACC}, an autoscaler that combines workload forecasting with SLO-budget feedback to reduce replica use while meeting fixed-period SLOs. BACC separates the base forecaster from its budget-aware scaling mechanism, allowing different forecasting models to be used without changing the budget-control logic. The base forecaster estimates future demand. BACC adjusts this estimate using recent forecast errors, with budget pacing setting the risk tolerance of this adjustment. A measured service profile then converts the adjusted demand into a replica target. A budget-aware trigger can request a replica increase before the next scheduled decision. BACC tracks the violation rate throughout the SLO period. It lowers its risk tolerance when violations accumulate too fast. It raises its risk tolerance when violations stay below the budget. Compared with fixed over-provisioning,
BACC uses fewer replicas. Compared with no over-provisioning, it still meets the SLO. Figure~\ref{fig:k8s_trace_d_dynamics} illustrates this behavior on a bursty Kubernetes workload.

We evaluate BACC through trace-driven simulation and real Kubernetes deployment. The simulation uses five Azure Functions traces and evaluates request-latency and CPU-threshold SLOs separately, each with 95\% and 99\% compliance targets. This tests whether the same budget-control approach supports both forms of SLO metrics. We compare BACC with reactive HPA and predictive OptScaler, using one BACC configuration for all settings.

In Kubernetes, we replay a regular and a bursty workload, running each autoscaler three times on each. HPA uses the same fixed utilization target on both workloads. On the regular workload, this target leaves most of the budget unused, with 1.94\% violations against a 5\% budget. On the bursty workload, the same target exceeds the budget, with 9.72\% violations. BACC meets the budget on both.

Our contributions are as follows:

\begin{itemize}
    \item We introduce \textbf{BACC}, a forecaster-independent autoscaler that treats a fixed-period SLO as a violation budget to spend. BACC over-provisions when little budget remains and accepts some violations when much remains. Its budget-control logic supports different SLO metrics through metric-specific replica sizing.

    \item We evaluate \textbf{BACC} with trace-driven simulation and Kubernetes replay. In simulation, BACC meets all 20 primary settings, compared with 16 for HPA and 12 for OptScaler. It uses up to 55.5\% fewer replicas where HPA also complies. In Kubernetes, BACC meets the SLO in all three runs on a bursty workload, whereas HPA meets it in none.

    \item Through component ablations in simulation, we show where the savings come from. Without over-provisioning, prediction alone fails in 10 of 20 settings. Fixed over-provisioning meets the SLO but leaves 42--100\% of the budget unused. Budget pacing reduces replicas by up to 35\% and still meets the SLO. Sensitivity experiments also examine the choice of forecaster.
\end{itemize}

\section{Related Work}
\label{sec:related}

\subsection{Reactive and Proactive Autoscaling}

We focus on horizontal autoscaling of containerized services, where capacity is adjusted by changing the number of replicas.

Reactive autoscalers adjust capacity in response to observed workload or service conditions. In a CPU-based configuration, Kubernetes Horizontal Pod Autoscaler (HPA)~\cite{kubernetes_hpa_docs} compares measured utilization with a target and adjusts replicas approximately according to their ratio. KEDA~\cite{keda_docs} supports event-driven signals such as queue depth. SATA~\cite{pozdniakova2024sla} adjusts HPA's utilization threshold according to observed service performance. These approaches respond to changing conditions, but additional replicas become available only after detection and startup.

Proactive autoscalers act before demand arrives so that capacity can become ready when needed. Kraken~\cite{bhasi2021kraken} forecasts demand for serverless workflows represented as directed acyclic graphs; OptScaler~\cite{zou2023optscaler} combines prediction, online capacity estimation, and model-predictive control with a fixed-confidence Gaussian chance constraint. AAPA~\cite{zhang2025aapa} uses workload patterns to select scaling policies, with confidence-based adjustments. These methods account for uncertainty, but their published decision rules do not explicitly adjust confidence settings according to the remaining fixed-period SLO allowance.

Prior resource managers improve efficiency through automatic resource configuration and performance-aware allocation across microservices~\cite{rzadca2020autopilot,qiu2020firm,zhang2021sinan,luo2022erms}. Other systems coordinate resource control and provisioning to meet application or workflow performance targets~\cite{wang2024autothrottle,zhou2022aquatope,sachidananda2024erlang}. BACC shares their objective of resource-efficient service operation, with a focus on horizontal replica provisioning. CloudScale~\cite{shen2011cloudscale} combines demand prediction with adaptive padding and reactive correction informed by resource pressure and SLO feedback. BACC additionally uses the violation allowance remaining in a fixed accounting period to set its risk tolerance.

\subsection{Online Conformal Calibration and SLO-Budget Control}

Site reliability engineering (SRE) practices use remaining error budgets and violation burn rates to guide operational decisions~\cite{beyer2016site,google_slo_monitoring}. BACC applies this accounting to replica provisioning. The remaining allowance indicates how many additional violations are affordable, while the burn rate indicates how quickly that allowance is being consumed. BACC uses these signals to set its risk tolerance.

BACC uses conformal calibration to adjust workload forecasts. Adaptive Conformal Inference (ACI)~\cite{gibbs2021adaptive} and Conformal PID~\cite{angelopoulos2023conformal} adapt calibration using prediction-coverage feedback. Their coverage target fixes the desired frequency of forecast misses, that is, a fixed risk tolerance; it does not by itself determine which target is appropriate for the service's remaining SLO allowance. Conformal Lyapunov Optimization~\cite{binucci2025conformal} combines online conformal risk control with long-term resource allocation, and Cost-Aware ACI~\cite{wu2026costaware} incorporates violation frequency and severity into runtime calibration. Both provide long-run guarantees and finite-time bounds under their respective assumptions.

BACC's distinction is its use of service violations and the allowance remaining before a predefined accounting period ends to set the risk tolerance for replica provisioning. The SLO requirement remains fixed, while the risk tolerance of individual provisioning decisions changes with budget use.

\section{Design}
\label{sec:design}

BACC chooses a risk tolerance from budget usage and provisions replicas for the forecast demand at that risk tolerance. Budget Pacing sets the risk tolerance from observed violations and the remaining budget. Forecast Calibration and Replica Sizing then convert the forecast into a replica target at this risk tolerance, and the Budget-Aware Trigger adds replicas early when violations run ahead of budget.

\subsection{System Model and Objective}
\label{sec:setting}

We consider a horizontally scalable containerized service whose SLO is evaluated over a fixed period of $T$ minutes. Requests are distributed among replicas with the same resource configuration, and BACC controls the number of provisioned replicas.

BACC measures workload and SLO violations every minute. Every $h$ minutes, it uses the measurements collected so far to decide how many replicas to provision. Newly requested replicas can serve requests only after their startup delay. Scale-downs take effect from the next minute.

During minute $t$, let $y_t$ denote the observed workload in requests per minute, $x_t$ the provisioned replica count, and $r_t$ the number of Ready replicas. The provisioned count includes both Ready replicas and replicas that are still starting; only Ready replicas can serve requests. Measurements for minute $t$ are collected at its end, before issuing a new replica decision.

We consider SLOs evaluated by counting requests or time intervals that meet a specified requirement. We call each counted request or interval an \emph{accounting unit}. The selected SLO defines the accounting unit and the condition under which it violates the requirement. Let $n_t$ denote the number of accounting units during minute $t$ and $b_t$ the number that violate the requirement. For request-based accounting, $n_t$ is the request count; for minute-based accounting, $n_t=1$.

BACC cannot control the workload. It controls only the number of replicas that serve it. This number affects violations through the load on each Ready replica. With sufficiently balanced requests and $r_t>0$, the per-replica load is approximately $y_t/(60r_t)$ requests/s. Fewer Ready replicas therefore mean a higher load per replica. A higher load increases queueing delay and CPU utilization, which raises the risk of violating a latency or CPU SLO. Section~\ref{sec:latency_translator} models this relationship with a measured service profile.

The SLO requires at least $1-\epsilon$ of the accounting units to comply over the period. The allowed fraction $\epsilon$ is the \emph{violation budget}, and the realized violation rate is
\begin{equation}
S_{vr}(T)
=
\frac{\sum_{t=1}^{T}b_t}
     {\sum_{t=1}^{T}n_t},
\end{equation}
and the SLO is satisfied when $S_{vr}(T)\le\epsilon$.

Resource cost is the mean provisioned replica count, $\bar R=(1/T)\sum_{t=1}^{T}x_t$, including pending replicas. BACC seeks successive replica decisions that minimize this cost while satisfying the fixed-period SLO:
\begin{equation}
\min \ \bar R
\qquad \text{subject to} \qquad S_{vr}(T)\le\epsilon.
\label{eq:autoscaling_objective}
\end{equation}

Equation~\eqref{eq:autoscaling_objective} constrains only the violation rate over the whole period. Individual decisions may therefore accept some violation risk. BACC uses this freedom to save replicas when the budget allows it.

\subsection{Design Overview}
\label{sec:design_overview}

The risk tolerance $a_k$ is the central variable of BACC. At regular decision $k$, BACC provisions for a calibrated demand over the next service window. We define $a_k$ as the target probability that the peak demand in this window exceeds the calibrated demand. A low $a_k$ yields a higher calibrated demand and over-provisions. A high $a_k$ yields a lower calibrated demand and saves replicas. In conformal prediction, $a_k$ is called the miscoverage level. The risk tolerance differs from the violation budget $\epsilon$. The budget limits SLO violations over the whole period, while $a_k$ limits forecast misses at a single decision. BACC can therefore set $a_k$ above $\epsilon$, as in Fig.~\ref{fig:k8s_trace_d_dynamics}.

Figure~\ref{fig:system_overview} shows how the components connect. The upper group provisions at a given risk tolerance (Section~\ref{sec:provisioning}). The base forecast flows through Forecast Calibration and Replica Sizing to a replica target. The lower group adapts the risk tolerance to budget use (Section~\ref{sec:budget_pacing}). Its output $a_k$ enters Forecast Calibration from below. Blue arrows act at regular decisions, every $h$ minutes. Orange dashed lines carry per-minute feedback. Each minute, BACC updates the service profile and scores completed forecast windows. The Budget-Aware Trigger (Section~\ref{sec:budget_aware_trigger}) can also request replicas before the next regular decision.

\begin{figure*}[htbp]
  \centering
  \resizebox{\textwidth}{!}{%
    \definecolor{baccblue}{HTML}{1678B5}
\definecolor{baccborder}{HTML}{55A8DA}
\definecolor{baccfill}{HTML}{E2F3FF}
\definecolor{bacctext}{HTML}{163B56}
\definecolor{feedbackorange}{HTML}{D97119}
\definecolor{externalborder}{HTML}{8A949E}
\definecolor{externalfill}{HTML}{F3F4F5}
\begin{tikzpicture}[
  font=\footnotesize, >=Stealth,
  regular/.style={->, line width=0.9pt, draw=baccblue},
  perminute/.style={->, line width=0.8pt, densely dashed, draw=feedbackorange},
  trunk/.style={line width=0.8pt, densely dashed, draw=feedbackorange},
  module/.style={draw=baccborder, rounded corners=4pt, fill=baccfill,
    text=bacctext, line width=0.9pt, align=center, inner sep=4pt,
    minimum width=2.6cm, minimum height=1.2cm, font=\footnotesize\bfseries},
  external/.style={draw=externalborder, fill=externalfill, text=bacctext,
    line width=0.8pt, align=center, inner sep=4pt},
  lab/.style={font=\scriptsize, align=center, inner sep=1.2pt}
]
\fill[baccfill!45, rounded corners=6pt] (-0.6,-0.45) rectangle (13.4,5.3);
\draw[draw=baccborder!70, rounded corners=6pt, line width=0.6pt] (-0.6,-0.45) rectangle (13.4,5.3);
\node[anchor=south west, text=baccblue, font=\footnotesize\bfseries] at (-0.5,-0.4) {BACC controller};

\node[module] (pred)  at (1.0,3.6) {Workload\\Prediction};
\node[module] (calib) at (6.1,3.6) {Forecast\\Calibration};
\node[module] (size)  at (11.2,3.6) {Replica\\Sizing};
\node[module] (pace)  at (6.1,1.0) {Budget\\Pacing};
\node[module] (trig)  at (11.2,1.0) {Budget-Aware\\Trigger};

\node[draw=baccblue, dashed, rounded corners=5pt, line width=0.7pt, inner sep=7pt,
  fit=(pred)(calib)(size)] (grpA) {};
\node[anchor=north west, text=baccblue, font=\scriptsize\bfseries, align=left, inner sep=2pt]
  at ([yshift=-2pt]grpA.south west) {Provisioning at a given\\risk tolerance (Sec.~\ref{sec:provisioning})};
\node[draw=baccblue, dashed, rounded corners=5pt, line width=0.7pt, inner sep=7pt,
  fit=(pace)] (grpB) {};
\node[anchor=north east, text=baccblue, font=\scriptsize\bfseries, align=right, inner sep=2pt]
  at ([xshift=-2pt]grpB.north west) {Adapting the\\risk tolerance to\\budget use (Sec.~\ref{sec:budget_pacing})};
\node[anchor=east, text=baccblue, font=\scriptsize\bfseries, inner sep=2pt]
  at ([xshift=-4pt]trig.west) {(Sec.~\ref{sec:budget_aware_trigger})};

\node[external, minimum width=3.3cm, minimum height=2.1cm] (svc) at (16.0,3.6) {};
\node[text=bacctext, font=\footnotesize\bfseries, align=center] at (16.0,4.25) {Managed service\\{\scriptsize\mdseries (Kubernetes Deployment)}};
\foreach \i in {0,1,2} {
  \node[draw=externalborder!140, fill=externalborder!70, minimum size=0.32cm, inner sep=0pt] at (15.1+0.45*\i,3.4) {};
}
\node[draw=externalborder!140, dashed, fill=white, minimum size=0.32cm, inner sep=0pt] at (16.45,3.4) {};
\node[font=\scriptsize, align=center] at (16.0,2.9) {Ready and pending pods};

\node[external, minimum width=3.3cm, minimum height=1.0cm, font=\footnotesize\bfseries] (mon) at (16.0,1.0) {Monitoring};
\draw[perminute] (svc.south) -- (mon.north);

\draw[regular] (pred.east) -- node[lab, above=2pt] {workload\\forecast} (calib.west);
\draw[regular] (calib.east) -- node[lab, above=2pt] {calibrated\\demand} (size.west);
\draw[regular] (size.east) -- node[lab, above=2pt] {replica\\target} (svc.west);
\draw[regular, line width=1.6pt] (pace.north) -- node[lab, right, font=\footnotesize\bfseries] {risk tolerance $a_k$} (calib.south);
\draw[perminute] (trig.north) -- node[lab, right] {early scale-up} (size.south);

\coordinate (trunkTop) at (18.1,5.9);
\coordinate (trunkBot) at (18.1,-1.0);
\draw[trunk] (mon.east) -- (18.1,1.0);
\draw[trunk] (trunkTop) -- (trunkBot);
\draw[trunk] (trunkTop) -- (1.0,5.9);
\draw[perminute] (1.0,5.9) -- node[lab, right, pos=0.5, xshift=1pt] {workload} (pred.north);
\draw[perminute] (6.1,5.9) -- node[lab, right, pos=0.5, xshift=1pt] {forecast errors} (calib.north);
\draw[perminute] (11.2,5.9) -- node[lab, right, pos=0.58, xshift=1pt] {per-replica load,\\SLO violations} (size.north);
\draw[trunk] (trunkBot) -- (6.1,-1.0);
\draw[perminute] (6.1,-1.0) -- node[lab, right, pos=0.72] {SLO violations} (pace.south);
\draw[perminute] (11.2,-1.0) -- node[lab, right, pos=0.72] {workload,\\SLO violations} (trig.south);

\draw[regular] (0.0,-1.6) -- ++(0.9,0) node[anchor=west, text=black, xshift=3pt] {regular decision (every few minutes)};
\draw[perminute] (6.6,-1.6) -- ++(0.9,0) node[anchor=west, text=black, xshift=3pt] {per-minute measurement or early action};
\end{tikzpicture}%
  }
  \caption{BACC overview. The five blue boxes are BACC's components; gray boxes show managed service and monitoring. At regular decisions, blue arrows carry control decisions. Orange dashed lines carry per-minute feedback and any sizing update from the Budget-Aware Trigger. Dashed groups mark provisioning at a given risk tolerance (Section~\ref{sec:provisioning}) and adapting the risk tolerance to budget use (Section~\ref{sec:budget_pacing}).}
  \label{fig:system_overview}
\end{figure*}

Algorithm~\ref{alg:bacc} gives the control loop. It records measurements in $\mathcal H$, recent forecast-error scores in $\mathcal S$, and issued plans in $\mathcal P$. Each plan stores its replica target, calibrated demand, risk tolerance, and pending replicas. Measurements are retained for the accounting period.

\begin{algorithm*}[t]
\caption{BACC control flow}
\label{alg:bacc}
\small
\DontPrintSemicolon
\SetCommentSty{textnormal}
\Input{SLO violation budget $\epsilon$; accounting period $T$;
regular decision interval $h$; replica startup delay $d$;
service profile $f$; initial workload history $\mathcal H_0$}
$\mathcal H\leftarrow\mathcal H_0$; $\mathcal S\leftarrow\varnothing$;
$\mathcal P\leftarrow\mathrm{InitialPlan}()$;
$a\leftarrow\epsilon$\;
\For{$t=1,\ldots,T$}{
  $o_t=(y_t,n_t,b_t,r_t,x_t)\leftarrow\mathrm{Monitor}()$
  \tcp*[r]{Collect the completed minute's measurements (\S\ref{sec:setting}).}
  $\mathcal H\leftarrow\mathrm{Append}(\mathcal H,o_t)$;
  $f\leftarrow\mathrm{UpdateProfile}(f,o_t)$
  \tcp*[r]{Update service risk from load and violations (\S\ref{sec:latency_translator}).}
  $\mathcal S\leftarrow\mathrm{UpdateCompletedScores}(\mathcal S,\mathcal H,\mathcal P,t)$
  \tcp*[r]{Score completed forecast windows (\S\ref{sec:forecast_calibration}).}
  \If{$\mathrm{RegularDecision}(t,h)$}{
    $\hat{\mathbf y}\leftarrow\mathrm{PredictWorkload}(\mathcal H,h+d)$
    \tcp*[r]{Forecast the next $h+d$ minutes (\S\ref{sec:prediction_module}).}
    $a\leftarrow\mathrm{BudgetPacing}(\mathcal H,\mathcal P)$
    \tcp*[r]{Select risk tolerance from feedback and remaining allowance (\S\ref{sec:budget_pacing}).}
    $(\tilde y^{\max},q)\leftarrow\mathrm{Calibrate}(\hat{\mathbf y},\mathcal S,a)$
    \tcp*[r]{Adjust demand with the selected error quantile (\S\ref{sec:forecast_calibration}).}
    $x^*\leftarrow\mathrm{SizeReplicas}(\tilde y^{\max},f)$
    \tcp*[r]{Divide demand by safe per-replica load (\S\ref{sec:capacity_planner}).}
    $x\leftarrow\mathrm{ApplyReplicaLimits}(x^*,\mathcal P)$
    \tcp*[r]{Apply replica limits (\S\ref{sec:capacity_planner}).}
    $\mathcal P\leftarrow\mathrm{IssuePlan}(\mathcal P,t,x,\hat{\mathbf y},q,a)$
    \tcp*[r]{Record target, forecast, calibration factor, and risk tolerance.}
  }
  \ElseIf{$\mathrm{WorkloadExceedsPlan}(\mathcal H,\mathcal P)
  \ \land\ \mathrm{AboveBudgetPace}(\mathcal H,\epsilon)$}{
    $x\leftarrow\mathrm{Replan}(\mathcal H,\mathcal P,\mathcal S,a,f)$
    \tcp*[r]{Recompute the target using the latest workload rise (\S\ref{sec:budget_aware_trigger}).}
    $\mathcal P\leftarrow\mathrm{IssueIncreaseAndHold}(\mathcal P,t,x)$
    \tcp*[r]{Request an increase and hold pending capacity until Ready (\S\ref{sec:budget_aware_trigger}).}
  }
}
\end{algorithm*}

\subsection{Provisioning at a Given Risk Tolerance}
\label{sec:provisioning}

\subsubsection{Workload Prediction}
\label{sec:prediction_module}

Workload Prediction produces the base forecast. It is the only demand input to provisioning; the next steps adjust it but do not replace it.

Regular decision $k$ occurs at the end of minute $t_k$, and $t_{k+1}=t_k+h$. The estimated startup delay $d$ includes scheduling, container startup, and service readiness. Replicas requested after minute $t_k$ spend the next $d$ minutes starting and first serve during minute $t_k+d+1$. Demand must therefore be predicted beyond the next decision.

At the regular decision after minute $t_k$, Workload Prediction uses observations through $t_k$ to produce one workload prediction for each of the next $h+d$ minutes, $\hat y_{t_k+j}$ for $j=1,\ldots,h+d$. Unlike the observed workload $y_t$, $\hat y_t$ denotes the predicted workload for minute $t$. The first $d$ predictions cover the startup minutes. During these minutes, newly requested replicas are still starting. Only replicas that are already Ready can serve requests. The remaining $h$ predictions cover the service window, when the new replicas are Ready. BACC provisions the same number of replicas for the whole window. Only the peak prediction in the window determines this number, because capacity for the peak also covers the other minutes. Figure~\ref{fig:decision_timeline} illustrates this timing.

BACC does not require a particular forecasting method. It needs only point forecasts; Forecast Calibration handles their uncertainty. The predictor may be a statistical model such as ARIMA~\cite{box2015time} or a learned model such as Chronos-2~\cite{ansari2025chronos2}, a recently proposed pretrained time-series foundation model from Amazon. Section~\ref{sec:main_comparison} evaluates both.

\begin{figure}[t]
  \centering
  \resizebox{\columnwidth}{!}{%
    \begin{tikzpicture}[x=0.60cm,y=0.55cm,font=\scriptsize]
  \foreach \i in {1,...,11} {
    \node at (\i-0.5,3.45) {\i};
  }
  \node[anchor=east] at (-0.1,3.45) {minute $t_k{+}$};
  \draw[gray!60,fill=orange!25] (0,1.6) rectangle (1,2.5);
  \draw[gray!60,fill=blue!15]   (1,1.6) rectangle (6,2.5);
  \node[anchor=east] at (-0.15,2.05) {decision $k$};
  \node at (0.5,2.9) {startup};
  \draw[gray] (0.5,2.67) -- (0.5,2.5);
  \node at (3.5,2.05) {sizing window};
  \draw[gray!60,fill=orange!25] (5,0) rectangle (6,0.9);
  \draw[gray!60,fill=blue!15]   (6,0) rectangle (11,0.9);
  \node[anchor=east] at (-0.15,0.45) {decision $k{+}1$};
  \node at (5.5,1.23) {startup};
  \node at (8.5,0.45) {sizing window};
  \draw[very thick] (0,-0.25) -- (0,2.7);
  \draw[very thick] (5,-0.25) -- (5,2.7);
  \node[below] at (0,-0.25) {end of $t_k$};
  \node[below] at (5,-0.25) {end of $t_{k+1}=t_k{+}h$};
\end{tikzpicture}%
  }
  \caption{Decision timing for $h=5$ and $d=1$. Decision $k$ follows the measurements for minute $t_k$ and forecasts minutes $t_k+1$ through $t_k+6$. Newly requested replicas start during $t_k+1$; the target is sized for minutes $t_k+2$ through $t_k+6$. The next regular decision occurs at the end of $t_k+5$.}
  \label{fig:decision_timeline}
\end{figure}

\subsubsection{Forecast Calibration}
\label{sec:forecast_calibration}

Forecast Calibration turns the base forecast into a calibrated demand. It uses two inputs: recent forecast errors and the risk tolerance $a_k$ from Section~\ref{sec:budget_pacing}. BACC measures how much past forecasts underpredicted or overpredicted demand. The risk tolerance then decides how much of this error to cover. A smaller $a_k$ covers more of the recent error and yields a higher calibrated demand. A larger $a_k$ covers less error and can yield a calibrated demand below the base forecast.

BACC calibrates the peak, i.e., the maximum, of the base forecast in the service window $W_k$, from minute $t_k+d+1$ through $t_k+d+h$:
{
\begin{equation}
\hat y_k^{\max}=\max_{t\in W_k}\hat y_t,
\qquad
y_k^{\max}=\max_{t\in W_k}y_t,
\label{eq:horizon_max_forecast}
\end{equation}%
}%
where $y_k^{\max}$ is known after the window completes. The ratio gives one relative \emph{nonconformity score} for the completed window:
\begin{equation}
s_k=\frac{y_k^{\max}+c}{\hat y_k^{\max}+c},
\label{eq:relative_score}
\end{equation}
where $c>0$ prevents division by zero and stabilizes the ratio at low workload. Scores above one indicate underprediction; scores below one indicate overprediction. The ratio allows errors at different workload scales to inform calibration. BACC clips each score to $[1/20,20]$ to limit the effect of extreme ratios. BACC keeps a history of recent scores. When the history is full, the oldest score is removed as a new score arrives.

Let the $m_k\ge1$ scores in the history be ordered as $s_{(1)}\le\cdots\le s_{(m_k)}$. Standard split conformal calibration selects the score at rank $\lceil(m_k+1)(1-a_k)\rceil$ \cite{vovk2005algorithmic,shafer2008tutorial,lei2018distribution}. For a small $a_k$, this rank can exceed $m_k$. BACC then uses the largest observed score:
\begin{equation}
j_k=\min\!\left\{m_k,
\left\lceil(m_k+1)(1-a_k)\right\rceil\right\},
\qquad q_k=s_{(j_k)}.
\label{eq:conformal_quantile}
\end{equation}
Thus, the largest observed score is the most conservative finite adjustment.

The selected score gives the calibrated demand:
\begin{equation}
\tilde y_k^{\max}
=
\max\!\left(0,(\hat y_k^{\max}+c)q_k-c\right).
\label{eq:conformal_adj}
\end{equation}
Coverage means $y_k^{\max}\le\tilde y_k^{\max}$, equivalently $s_k\le q_k$. Conformal calibration assumes that the next forecast error behaves like the recent ones. Under this assumption, the calibrated demand is exceeded in about a fraction $a_k$ of windows. Real workloads change over time, so the next error can be larger or smaller than the recent ones. The actual miss rate can therefore differ from $a_k$. Section~\ref{sec:budget_pacing} compensates by adjusting $a_k$ from observed violations. Before any forecast window has completed, BACC sets $q_k=1$. Replica Sizing uses $\tilde y_k^{\max}$ to choose the target.

\subsubsection{Replica Sizing}
\label{sec:translating_replicas}

Replica Sizing converts the calibrated demand into a replica count. It needs one quantity: the highest load that one replica can serve while its violation risk stays within the budget.

\paragraph{Estimating Service Risk from Load}
\label{sec:latency_translator}

Replica Sizing uses a measured profile $f(\lambda)$ of violation probability at per-replica offered load $\lambda$ requests/s. The profile follows the selected SLO's accounting rule. It is initialized from load tests that measure the service's violations at different offered loads before BACC begins operation.

BACC stores violation and accounting-unit counts on a fixed grid of profiled loads. It fits $f$ as a nondecreasing function of load. The pool-adjacent-violators algorithm (PAVA) pools adjacent points whose risks decrease as load increases. It combines their counts until the estimates are nondecreasing. BACC then selects the highest profiled load whose estimated risk is within the budget:
\begin{equation}
\lambda_{\mathrm{safe}}
=
\max\{\lambda:f(\lambda)\le\epsilon\}.
\label{eq:safe_load}
\end{equation}
If no profiled load satisfies $f(\lambda)\le\epsilon$, BACC falls back to maximum-capacity sizing, subject to the scaling limits below.

Runtime observations update the nearest grid point. BACC refits the curve every five minutes with usable load and SLO measurements. The refit interval is configurable. Section~\ref{sec:cpu_eval} specifies the estimator used for each SLO.

Sizing uses the fixed budget $\epsilon$, not the risk tolerance $a_k$. The risk tolerance acts only on the calibrated demand. The risk limit $\epsilon$ and the SLO threshold stay fixed.

Profiles assume representative deployment conditions and sufficiently balanced requests across replicas. Sizing uses only the measured load range. Changes in hardware, container limits, request mix, or dependencies may require new measurements.

\paragraph{Choosing the Replica Count}
\label{sec:startup_delay}
\label{sec:capacity_planner}

Given the calibrated demand and the load one replica can serve, BACC divides the former by the latter and rounds up:
\begin{equation}
x_k^*
=
\left\lceil
\frac{\tilde y_k^{\max}/60}{\lambda_{\mathrm{safe}}}
\right\rceil.
\label{eq:replica_translation}
\end{equation}
The factor of 60 converts requests per minute to requests per second. BACC then applies the deployment's replica bounds and maximum change per decision. Changes are measured from the \emph{planned} count, including Ready and pending replicas, to avoid requesting the same capacity twice or unintentionally canceling pending scale-ups. A scale-down takes effect in the next minute.

\subsection{Adapting the Risk Tolerance to Budget Use}
\label{sec:budget_pacing}

Budget Pacing first adjusts the risk tolerance from observed violations, then applies a remaining-budget ceiling. Its output $a_k$ selects the calibration quantile used for replica sizing.

\subsubsection{Adjusting the Risk Tolerance from Violation Feedback}

At regular decision $k$, BACC computes the rolling violation rate $\hat v_k^{\mathrm{rol}}$ and the rate since the accounting period began, $\hat v_k^{\mathrm{cum}}$. Each divides violations by accounting units in its window. The rolling window covers the latest $w$ minutes, or all elapsed minutes if fewer are available. It detects recent deterioration, while the cumulative rate retains earlier violations. The pacing error is $e_k=\max(\hat v_k^{\mathrm{rol}},\hat v_k^{\mathrm{cum}})-\epsilon$; positive error calls for a lower risk tolerance.

Following Adaptive Conformal Inference~\cite{gibbs2021adaptive}, but using SLO-pacing error rather than coverage feedback, BACC updates from $a_0^{\mathrm{pace}}=\epsilon$:
\begin{equation}
\begin{aligned}
a_k^{\mathrm{pace}}
&=
\mathrm{clip}\!\left(
a_{k-1}^{\mathrm{pace}}-\eta_k e_k,
a_{\mathrm{pace,min}},a_{\mathrm{pace,max}}
\right),\\
\eta_k
&=
\begin{cases}
\eta_{\mathrm{tighten}}, & e_k>0,\\
\eta_{\mathrm{relax}}, & e_k\le0.
\end{cases}
\end{aligned}
\label{eq:protection_update}
\end{equation}
BACC clips $a_k^{\mathrm{pace}}$ to $[a_{\mathrm{pace,min}},a_{\mathrm{pace,max}}]$, with both bounds in $[0,1]$.

\subsubsection{Protecting the Remaining Budget}
\label{sec:conditional_risk}

BACC next calculates the highest risk tolerance the remaining budget can afford, which we call the ceiling. The feedback update reacts only after violations occur. When little budget remains, a few more violations can exhaust it. The ceiling lowers $a_k$ before this happens. In Fig.~\ref{fig:k8s_trace_d_dynamics}, it lowers the risk tolerance from 31.7\% to 12.8\% at the late burst. The ceiling limits only the current $a_k$; it does not change the pacing state. BACC computes the ceiling in three steps. All counts below are cumulative over the accounting period up to decision $k$; we omit $k$ from $K_i$, $U_i$, $p_0$, and $p_1$.

\emph{Step 1: affordable violation rate.} This step asks how many violations the remaining accounting units can still afford. After $t_k$ minutes, the period has $N_k$ units and $B_k$ violations. BACC estimates the remaining units as $N_k^{\rm rem}=(N_k/t_k)(T-t_k)$, assuming the observed average rate continues. It also reserves $\rho_k=\hat v_k^{\rm last}\min(dN_k/t_k,N_k^{\rm rem})$ violations for the startup minutes, when new replicas cannot yet help. Here $\hat v_k^{\rm last}=b_{t_k}/n_{t_k}$ is the violation fraction of the latest minute, or 0 if $n_{t_k}=0$. The affordable violation rate divides the allowance left after these violations by the remaining units:
{
\begin{equation}
u_k=\mathrm{clip}\!\left(
\frac{\epsilon(N_k+N_k^{\rm rem})-B_k-\rho_k}{N_k^{\rm rem}},0,1\right).
\label{eq:remaining_request_risk}
\end{equation}%
}%
For a minute-based SLO, $N_k^{\rm rem}=T-t_k$ is known, the allowance is $\lfloor\epsilon T\rfloor-B_k$, and the reserve is up to one violation for each of the next $\min(d,N_k^{\rm rem})$ minutes. The ceiling applies only while units remain.

\emph{Step 2: affordable miss probability.} This step asks how often the forecast can be missed at the rate $u_k$. A unit has a forecast miss when workload in its minute exceeds the calibrated demand of its plan. Let $p_1$ and $p_0$ be the violation probabilities of units with and without a miss. For $K_i$ violations among $U_i$ units in case $i\in\{0,1\}$, BACC uses $p_i=(K_i+1/2)/(U_i+1)$, the posterior mean under the Jeffreys $\operatorname{Beta}(1/2,1/2)$ prior~\cite{jeffreys1946invariant}, which avoids estimates of exactly 0 or 1. If misses occur with probability $p_F$, the expected violation rate is $p_Fp_1+(1-p_F)p_0$. When $u_k<\epsilon$ and $p_1>p_0$, BACC sets this rate to $u_k$ and solves for $p_F$:
{
\begin{equation}
p_F^{\max}
=
\mathrm{clip}\!\left(
\frac{u_k-p_0}{p_1-p_0},0,1
\right).
\label{eq:unit_forecast_risk_ceiling}
\end{equation}%
}%
Otherwise, $p_F^{\max}=1$, which places no limit on forecast misses. If $u_k<p_0$, the budget cannot afford even the violations without misses, so $p_F^{\max}=0$.

\emph{Step 3: conversion to a risk tolerance.} The value $p_F^{\max}$ is a miss probability per accounting unit, whereas the risk tolerance is a miss probability per service window. A missed window may exceed its calibrated demand in only some of its minutes, which contain different numbers of units. BACC therefore compares $M_k$, the units in minutes that exceeded their calibrated demand, with $A_k$, the sum of recorded risk tolerances over units with assigned plans. Each such minute contributes $n_ta$ to $A_k$, where $a$ is the risk tolerance recorded with its plan. The conversion factor is $\kappa_k=\mathrm{clip}((M_k+1/2)/(A_k+1/2),0.1,10)$; the bounds limit small-sample effects. The ceiling is $p_F^{\max}/\kappa_k$, and the final risk tolerance is
\begin{equation}
a_k
=
\max\!\left(
a_{\mathrm{cal,min}},
\min\!\left(
a_k^{\mathrm{pace}},
\frac{p_F^{\max}}{\kappa_k}
\right)
\right).
\label{eq:remaining_budget_ceiling}
\end{equation}
The lower bound $a_{\mathrm{cal,min}}>0$ is set below $a_{\mathrm{pace,min}}$, so the ceiling can request a lower risk tolerance than pacing alone. A lower $a_k$ selects a higher score quantile and can increase the replica target. These empirical estimates guide budget protection but do not guarantee compliance for every workload.

\subsection{Budget-Aware Trigger}
\label{sec:budget_aware_trigger}

To avoid waiting up to $h$ minutes for the next decision, the trigger checks each minute whether the workload exceeds the workload used to size the current plan and whether any of the latest, rolling, or cumulative violation rates exceeds $\epsilon$. An early update reuses the latest forecast and calibration and pacing state, without advancing their regular schedule or adding a score. It sets a minimum sizing workload $y_t^{\mathrm{early}}=y_t+(1+d)\max(y_t-y_{t-1},0)$. This extends a positive workload increase over $1+d$ minutes. The update can only increase replicas, which still incur startup delay. BACC holds the pending target until the replicas become Ready.

\section{Evaluation}
\label{sec:evaluation}

We organize the evaluation around five research questions:
\begin{enumerate}[\renewcommand{\labelenumi}{\textbf{RQ\arabic{enumi}.}}%
  \IEEEsetlabelwidth{\textbf{RQ5.}}\setlength{\itemsep}{2pt}]
\item \textbf{Compliance and resource use:} Does BACC meet fixed-period request-latency SLOs while using fewer replicas than reactive and predictive baselines?
\item \textbf{SLO metric adaptation:} Does the same budget-pacing loop support a CPU-threshold SLO when only the replica-sizing component changes?
\item \textbf{Startup-delay} robustness: How does BACC perform when scale-up takes longer than the controller assumes?
\item \textbf{Forecaster sensitivity:} How does changing the forecaster affect BACC's compliance and replica use?
\item \textbf{Component contributions:} How do BACC's individual components contribute to compliance and resource efficiency?
\end{enumerate}

RQ1 and RQ2 test whether BACC meets the SLO with fewer replicas, RQ3 and RQ4 test its robustness, and RQ5 isolates the effect of adapting the risk tolerance to budget use. We answer these questions using a trace-driven simulator built from measurements of the Kubernetes service described in Section~\ref{sec:generality}.

\subsection{Experimental Setup}
\label{sec:cpu_eval}

\paragraph{Workload selection}
We use five functions from the Azure Functions Trace 2019~\cite{shahrad2020serverless}, each sampled for 14 days at one-minute granularity. This stratified case-study set spans distinct operating regimes; it is not a random sample for population-wide inference. The extraction artifact contains full anonymized identifiers, with suffixes in Table~\ref{tab:trace_stats}. A is smooth and periodic, B is a regular HTTP workload, C has almost no daily correlation, D is most bursty and spectrally complex, and E has the largest mean demand. The set spans event, HTTP, and queue triggers and more than $15\times$ in mean demand.

\begin{figure*}[t]
    \centering
    \includegraphics[width=0.94\textwidth]{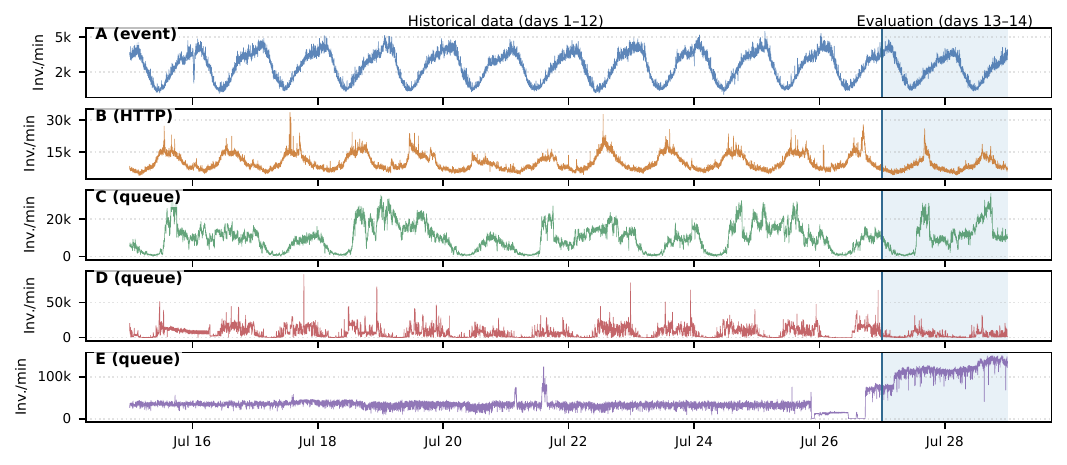}
    \caption{The five Azure Functions traces and their chronological split. Each trace uses its own vertical scale so that both smooth variation and bursts remain visible. Days~1--12 provide historical data, and days~13--14 are reserved for evaluation.}
    \label{fig:workload_traces}
\end{figure*}

Figure~\ref{fig:workload_traces} shows the complete traces and chronological split. Days~1--12 are the historical prefix, with days~11--12 used to select one HPA configuration shared by all traces. Controllers run for all 2,880 minutes of days~13--14. For every method, we exclude the first 5 minutes as a common warm-up when calculating the reported SLO, resource, and scaling-churn metrics, leaving 2,875 measured minutes. Cumulative budget state is not reset between the two days. The 1,440-minute feedback window is a rolling responsiveness signal within this 2,880-minute period. The predictive methods initialize their forecast models from the historical prefix and update them online during evaluation; no method observes a future request count. We replay the native request counts with uniform spacing inside each minute.

\begin{table}[t]
    \centering
    \caption{Azure Functions workloads (invocations/min). ID gives the final
    four hexadecimal digits of the anonymized function identifier. CV and
    P99/P50 summarize burstiness, AC(1d) is the one-day autocorrelation,
    Entropy is normalized spectral entropy, and Zero\% is the fraction of
    minutes with no invocations.}
    \label{tab:trace_stats}
    \resizebox{\columnwidth}{!}{%
\begin{tabular}{lllrrrrrr}
\toprule
Trace & ID & Trigger & Mean & CV & P99/P50 & AC(1d) & Entropy & Zero\% \\
\midrule
A & 4b3e & event & 2868 & 0.36 & 1.53 & 0.89 & 0.10 & 0.00 \\
B & b559 & HTTP & 10124 & 0.35 & 2.18 & 0.74 & 0.23 & 0.00 \\
C & 1af2 & queue & 9920 & 0.69 & 2.85 & 0.06 & 0.37 & 0.04 \\
D & 6486 & queue & 6734 & 1.08 & 6.38 & 0.44 & 0.55 & 0.64 \\
E & 1d5c & queue & 44679 & 0.70 & 4.13 & 0.39 & 0.28 & 2.06 \\
\bottomrule
\end{tabular}

    }
    \end{table}

\paragraph{Simulation environment}
We model service behavior from Kubernetes measurements. The containerized service in Section~\ref{sec:generality} has a CPU request and limit of 250~mCPU. Independent load-test batches profile one and eight replicas and record offered load, request outcomes, latency, and application-container CPU. The controller-visible profile covers 2--28 requests/s/pod. Separate simulator-fit and validation profiles extend to 100 requests/s/pod, so the environment covers overload that the controllers did not observe during calibration.

We fit monotone curves against per-pod offered load. BACC's service profile and OptScaler's CPU model see the controller-calibration curve; the hidden simulator-fit curve generates outcomes, and the third curve independently validates the fit. Across 24 common load points, the mean absolute differences are 0.032 percentage points in request-violation probability, 1.84 percentage points in CPU-violation probability, and 5.8~mCPU in CPU use per pod.

As a stability check, we replace the hidden simulator-fit profile with the independently collected validation profile without retuning BACC. BACC remains compliant in all 20 combinations; mean provisioned replicas change by at most 0.85\% in any metric/target setting, and the largest absolute violation-rate change is 0.104 percentage points.

For latency, a request counts as a violation when it is non-2xx or slower than 500~ms; the number of violating requests in each minute is sampled from the hidden empirical probability. For CPU, the simulator samples from the hidden empirical utilization distribution at the offered load. We use seed~42 when sampling request violations and CPU utilization. We add neither synthetic measurement noise nor capacity drift. The simulator advances in one-minute steps. A scale-up incurs a one-step startup delay before the new replicas can serve traffic, whereas a scale-down takes effect without an actuation delay.

\paragraph{SLOs and metrics}
Latency and CPU are separate SLO instantiations; their violations are not combined. Latency is the request duration measured at the managed service's destination-side proxy, rather than client-observed end-to-end latency across multiple services. A request that is both non-2xx and slower than 500~ms is counted once. For latency, $S_{vr}$ is the number of violating requests divided by all offered requests. For CPU, a minute violates when the mean application-container CPU utilization across all pods exceeds $c^*=0.5$, and $S_{vr}$ is the fraction of violating minutes. CPU utilization is relative to the application container's CPU request; each minute contributes one accounting unit. The 95\% and 99\% compliance targets therefore require $S_{vr}\le5\%$ and $S_{vr}\le1\%$, respectively. Resource use is mean provisioned replicas $\bar R$, including pending replicas; service outcomes depend only on Ready capacity. We call a resource difference a saving only when both compared methods meet the requested SLO.

BACC's latency profile uses the fraction of bad requests among all requests at each pooled load point. Its CPU profile counts one accounting unit per minute and uses a one-sided 95\% upper confidence bound on violation probability. For pooled counts of $K$ violations among $U$ accounting units, this bound is the 0.95 quantile of $\operatorname{Beta}(K+1,U-K)$, set to one when $K=U$. The upper bound avoids treating a load as safe simply because few violations were observed in a small sample.

All controllers start with the same replica count within each (trace, metric, target) setting. We choose this count using the workload during the last hour of day~12 and the controller-visible service profile.

\paragraph{Baselines}
We compare against reactive HPA and predictive OptScaler.

\textbf{HPA}~\cite{kubernetes_hpa_docs} uses the standard CPU-utilization ratio rule and the official scale-up policy, which permits at most the larger of a 100\% increase and four pods per decision. On days~11--12, we jointly grid-search its CPU target and scale-down stabilization window for each (metric instantiation, compliance level). CPU-target candidates range from 0.20 to 1.00 in increments of 0.05, and scale-down windows are 0, 3, and 5 minutes. Each candidate is evaluated with seed~42. A setting is shared by all five traces and is eligible only when every trace meets the calibration budget; among eligible settings, we select the one with the lowest equal-trace mean provisioned replicas. All four selected settings use a 5-minute scale-down stabilization window. Their CPU targets are 0.55/0.30 for the latency 95\%/99\% targets and 0.35/0.20 for the CPU 95\%/99\% targets.

\textbf{OptScaler}~\cite{zou2023optscaler} has no public implementation, so we implement its CPU-model MPC from the description in the paper, retaining an optimization horizon of $D=11$ decision intervals and setting its fixed chance confidence $\alpha$ to the requested compliance target, 0.95 or 0.99. Its original workload predictor is also not publicly available. We therefore give OptScaler the same workload predictor as BACC. In both panels, it scales from CPU, as in the original design; only the metric used to score its trajectory changes.

\paragraph{Configuration}
Table~\ref{tab:controller_config} lists the settings. BACC and OptScaler share AutoARIMA with a 720-minute context and periodic stepwise AIC selection of the ARIMA orders~\cite{hyndman2008automatic}, isolating their conversion of forecasts and service observations into replica decisions. No settings are trace-specific.

\begin{table*}[!t]
\centering
\caption{Autoscaler configurations in the primary comparison. HPA settings are selected on days~11--12 and shared by all traces.}
\label{tab:controller_config}
\small
\setlength{\tabcolsep}{3pt}
\begin{tabular}{@{}>{\raggedright\arraybackslash}p{0.12\textwidth}
                    >{\raggedright\arraybackslash}p{0.25\textwidth}
                    >{\raggedright\arraybackslash}p{0.58\textwidth}@{}}
\toprule
Method & Parameter & Value \\
\midrule
Common & Evaluation warm-up & First 5 minutes \\
 & Initial replicas & Latency 95\%: A--E = 3/6/9/12/49; latency 99\%: 4/8/12/16/69; CPU 95\%: 5/8/13/18/74; CPU 99\%: 5/9/13/18/77 \\
\addlinespace
HPA & Decision interval & 1 minute \\
 & Scale-up policy & $\max(100\%,4\text{ replicas})$ per decision \\
 & Scale-down stabilization & 5 minutes \\
\addlinespace
OptScaler & Workload predictor & AutoARIMA; 720-minute context \\
 & Decision interval & 5 minutes \\
 & MPC horizon & $D=11$ \\
 & Chance confidence & $\alpha=0.95/0.99$ \\
 & Maximum change & 100 replicas per decision \\
\addlinespace
BACC & Workload predictor & AutoARIMA; 720-minute context \\
 & Decision interval & 5 minutes \\
 & Maximum change & 100 replicas per decision \\
 & Budget-feedback window & 1,440 minutes \\
 & Pacing bounds & $[a_{\mathrm{pace,min}},a_{\mathrm{pace,max}}]=[0.2\epsilon,0.9]$ \\
 & Pacing step sizes & $\eta_{\mathrm{tighten}}=1.0$, $\eta_{\mathrm{relax}}=0.2$ \\
 & Risk-tolerance lower bound & $a_{\mathrm{cal,min}}=0.1\epsilon$ \\
\bottomrule
\end{tabular}
\end{table*}

BACC's configuration is shared across traces, metrics, and compliance targets: a five-minute decision interval, one-minute startup delay, 12-hour calibration history, and 24-hour feedback window. Target-dependent bounds scale with $\epsilon$, and $T$ follows the accounting period. No choice is selected from days~13--14.

\subsection{Primary Results}
\label{sec:main_comparison}

\paragraph{RQ1: Request-latency SLO}
\begin{table}[!t]
\centering
\caption{Per-trace request-latency results (seed 42). Each cell reports $S_{vr}$ (\%) / mean provisioned replicas $\bar R$; bold violation rates exceed the 5\% or 1\% budget.}
\label{tab:primary_latency}
\scriptsize
\setlength{\tabcolsep}{3pt}
\begin{tabular}{lccc}
\toprule
Trace & HPA & OptScaler & BACC (ours) \\
\midrule
\multicolumn{4}{l}{\textbf{95\% compliance target}: $S_{vr}$ / $\bar R$} \\
\midrule
A & 0.57 / 2.7 & 0.10 / 3.1 & 1.34 / 2.4 \\
B & 1.16 / 7.4 & 0.28 / 8.8 & 2.74 / 6.4 \\
C & 1.28 / 9.3 & 0.48 / 10.8 & 3.07 / 8.1 \\
D & \textbf{9.96} / 4.3 & \textbf{8.18} / 4.6 & 3.95 / 6.8 \\
E & 1.28 / 94.6 & 0.43 / 109.1 & 3.91 / 76.1 \\
\midrule
\multicolumn{4}{l}{\textbf{99\% compliance target}: $S_{vr}$ / $\bar R$} \\
\midrule
A & 0.00 / 4.4 & 0.05 / 3.2 & 0.14 / 3.0 \\
B & 0.00 / 13.6 & 0.16 / 9.3 & 0.35 / 8.9 \\
C & 0.01 / 17.0 & 0.31 / 11.5 & 0.48 / 11.2 \\
D & \textbf{3.10} / 7.4 & \textbf{7.63} / 4.8 & 0.78 / 8.2 \\
E & 0.00 / 175.8 & 0.19 / 116.9 & 0.63 / 106.8 \\
\bottomrule
\end{tabular}
\end{table}

Table~\ref{tab:primary_latency} shows that BACC meets all ten latency settings. HPA and OptScaler each meet eight: both fail the 95\% and 99\% targets on bursty trace~D. BACC lowers D's violation rate from 9.96\% (HPA) and 8.18\% (OptScaler) to 3.95\% at the 95\% target, and from 3.10\% and 7.63\% to 0.78\% at the 99\% target.

On the four traces where HPA and BACC both comply, BACC saves 10.7--19.6\% per trace at the 95\% target and 31.2--39.3\% at the 99\% target. In cases where OptScaler complies, the corresponding ranges are 22.8--30.3\% and 2.1--8.6\%. OptScaler's violation rates and replica use change modestly when its fixed-chance confidence increases from 0.95 to 0.99. BACC lowers its risk tolerance as the violation budget tightens, narrowing the resource gap to OptScaler while retaining compliance on D.

\paragraph{RQ2: CPU-threshold SLO}
\begin{table}[!t]
\centering
\caption{Per-trace CPU-threshold results (seed 42). Each cell reports $S_{vr}$ (\%) / mean provisioned replicas $\bar R$; bold violation rates exceed the 5\% or 1\% budget.}
\label{tab:primary_cpu}
\scriptsize
\setlength{\tabcolsep}{3pt}
\begin{tabular}{lccc}
\toprule
Trace & HPA & OptScaler & BACC (ours) \\
\midrule
\multicolumn{4}{l}{\textbf{95\% compliance target}: $S_{vr}$ / $\bar R$} \\
\midrule
A & 0.00 / 3.8 & 1.04 / 3.1 & 0.77 / 3.1 \\
B & 0.07 / 11.7 & 4.66 / 8.8 & 3.51 / 9.3 \\
C & 0.49 / 14.7 & \textbf{8.90} / 10.8 & 3.41 / 12.7 \\
D & \textbf{6.64} / 6.4 & \textbf{15.10} / 4.6 & 4.87 / 7.0 \\
E & 0.00 / 148.8 & \textbf{8.21} / 109.1 & 4.17 / 112.3 \\
\midrule
\multicolumn{4}{l}{\textbf{99\% compliance target}: $S_{vr}$ / $\bar R$} \\
\midrule
A & 0.00 / 6.5 & 0.31 / 3.2 & 0.24 / 3.2 \\
B & 0.00 / 20.6 & \textbf{1.50} / 9.3 & 0.42 / 9.9 \\
C & 0.00 / 25.8 & \textbf{4.70} / 11.5 & 0.73 / 15.0 \\
D & \textbf{2.61} / 10.8 & \textbf{13.74} / 4.8 & 0.83 / 12.1 \\
E & 0.00 / 265.8 & 0.97 / 116.9 & 0.63 / 118.3 \\
\bottomrule
\end{tabular}
\end{table}

Table~\ref{tab:primary_cpu} gives the same comparison under minute-level CPU accounting. BACC again meets all ten settings. HPA meets eight and fails only trace~D; OptScaler meets four. On D, BACC reduces HPA's 6.64\%/2.61\% to 4.87\%/0.83\% at the 95\%/99\% targets. OptScaler's fixed-confidence CPU model yields non-compliant results for C--E at the 95\% target and B--D at the 99\% target, even though several of those trajectories satisfy the request-weighted latency SLO in Table~\ref{tab:primary_latency}.

For the four traces jointly compliant with HPA, BACC saves 13.5--24.5\% per trace at the 95\% target and 41.9--55.5\% at the 99\% target. Compared with OptScaler, BACC uses nearly the same resources on trace~A at both targets, 6.2\% more on B at the 95\% target, and 1.2\% more on E at the 99\% target. BACC meets all six remaining cases where OptScaler fails, supporting both metrics through the shared loop and metric-specific Replica Sizing.

\paragraph{Repeated-seed stability}
With forecasts, configurations, traces, and profiles fixed, we vary only hidden service-outcome draws using seeds 40--49. BACC meets all 200 cells, versus 160/200 for HPA and 105/200 for OptScaler (Table~\ref{tab:seed_stability}). HPA and OptScaler fail all 40 D cells, which BACC meets; OptScaler is additionally sensitive to draws under the CPU SLO.

On traces A--C and E, BACC and HPA jointly comply in all 160 paired cells, and BACC uses fewer replicas in every cell. Across the four trace-level paired means, its savings are 10.6--19.7\% and 30.6--39.3\% for latency at the 95\% and 99\% targets, and 11.7--24.0\% and 44.0--54.9\% for CPU. The corresponding per-trace 95\% confidence intervals all exclude zero. Against OptScaler, BACC likewise uses fewer replicas in all 80 jointly compliant latency cells on A--C and E. The confidence intervals quantify variation across service-profile draws, not variation across workloads.

\begin{table}[!t]
\centering
\caption{Stability of all methods across service-outcome seeds 40--49.
``Met'' counts compliant (trace, seed) cells. Means are computed across ten
seed-level equal-trace means; Max is the largest per-cell violation rate.
All 95\% confidence-interval half-widths are at most
0.15 percentage points for $S_{vr}$ and 0.31 replicas for $\bar R$.}
\label{tab:seed_stability}
\scriptsize
\setlength{\tabcolsep}{3pt}
\begin{tabular}{lllccc}
\toprule
SLO & Metric & Method & Met & Mean / Max $S_{vr}$ (\%) & Mean $\bar R$ \\
\midrule
\multirow{6}{*}{95\%}
 & \multirow{3}{*}{Latency}
   & HPA       & 40/50 & 2.78 / \textbf{10.44} & 23.67 \\
 & & OptScaler & 40/50 & 1.99 / \textbf{9.07}  & 27.22 \\
 & & BACC      & 50/50 & 3.03 / 4.18           & 19.93 \\
\cmidrule(lr){2-6}
 & \multirow{3}{*}{CPU}
   & HPA       & 40/50 & 1.40 / \textbf{6.82}  & 37.18 \\
 & & OptScaler & 11/50 & 8.09 / \textbf{15.83} & 27.22 \\
 & & BACC      & 50/50 & 3.25 / 4.90           & 29.35 \\
\midrule
\multirow{6}{*}{99\%}
 & \multirow{3}{*}{Latency}
   & HPA       & 40/50 & 0.60 / \textbf{3.15}  & 43.71 \\
 & & OptScaler & 40/50 & 1.68 / \textbf{7.94}  & 29.07 \\
 & & BACC      & 50/50 & 0.47 / 0.86           & 27.76 \\
\cmidrule(lr){2-6}
 & \multirow{3}{*}{CPU}
   & HPA       & 40/50 & 0.53 / \textbf{2.92}  & 66.09 \\
 & & OptScaler & 14/50 & 4.41 / \textbf{14.26} & 29.07 \\
 & & BACC      & 50/50 & 0.60 / 0.87           & 32.10 \\
\bottomrule
\end{tabular}
\end{table}

\paragraph{RQ3: Startup-delay mismatch}
We repeat all 60 seed-42 cells with realized startup delay increased from one to two minutes. BACC and OptScaler retain their one-minute assumption; all other inputs and HPA settings remain fixed. BACC complies in all 20 settings. Across the four metric/target groups, the equal-trace mean provisioned replicas increase by 1.1--4.6\%, and the largest absolute per-trace change in violation rate is 0.345 percentage points. Under the same mismatch, HPA remains compliant in 16/20 settings, while OptScaler's compliance count decreases from 12/20 to 10/20.

\paragraph{RQ4: Forecaster sensitivity}
Replacing AutoARIMA with Chronos-2 central point forecasts~\cite{ansari2025chronos2}, with all other inputs and seed~42 fixed, preserves compliance in all 20 settings. Mean replicas change by at most 1.8\% on A--C and E, with larger differences on D. This comparison demonstrates that BACC can use either forecaster with the same budget-control logic, although the resulting resource use remains sensitive to the forecasts.

On trace~D, Chronos-2 yields lower calibrated peak-demand estimates than AutoARIMA for latency and higher estimates for CPU, so BACC provisions fewer replicas for latency (5.7 vs.\ 6.8 at the 95\% target) and more for CPU (9.1 vs.\ 7.0 and 13.7 vs.\ 12.1 at the 95\% and 99\% targets).

\subsection{RQ5: Component Ablation}
\label{sec:ablation}

\begin{table*}[!t]
\centering
\caption{Incremental component ablation (seed 42). Each cell reports $S_{vr}$
(\%) / mean provisioned replicas $\bar R$; bold violation rates exceed the
requested budget. Replica Sizing is present in every variant.}
\label{tab:unified_ablation}
\scriptsize
\setlength{\tabcolsep}{3.5pt}
\begin{tabular}{llccccc}
\toprule
SLO & Variant & A & B & C & D & E \\
\midrule
\multicolumn{7}{l}{\textbf{95\% compliance target}: $S_{vr}$ / $\bar R$} \\
\midrule
\multirow{4}{*}{Latency}
 & Workload Prediction + Replica Sizing & 1.42/2.4 & 3.36/5.9 & 3.95/7.5 & \textbf{18.54}/3.4 & 4.82/71.4 \\
 & $+$ Forecast Calibration & 0.80/2.5 & 1.35/7.1 & 1.27/9.3 & 1.39/7.7 & 2.88/79.4 \\
 & $+$ Budget Pacing & 1.42/2.4 & 3.36/6.0 & 4.12/7.4 & 3.72/7.7 & 4.49/72.7 \\
 & Full BACC & 1.34/2.4 & 2.74/6.4 & 3.07/8.1 & 3.95/6.8 & 3.91/76.1 \\
\addlinespace
\multirow{4}{*}{CPU}
 & Workload Prediction + Replica Sizing & 1.32/3.0 & \textbf{7.37}/8.6 & \textbf{10.92}/10.6 & \textbf{15.41}/4.5 & \textbf{11.34}/107.8 \\
 & $+$ Forecast Calibration & 0.00/3.4 & 0.24/10.4 & 0.49/13.7 & 1.08/10.9 & 0.28/119.8 \\
 & $+$ Budget Pacing & 1.18/3.1 & 3.37/9.2 & 3.48/13.1 & 4.83/9.0 & 3.97/111.9 \\
 & Full BACC & 0.77/3.1 & 3.51/9.3 & 3.41/12.7 & 4.87/7.0 & 4.17/112.3 \\
\midrule
\multicolumn{7}{l}{\textbf{99\% compliance target}: $S_{vr}$ / $\bar R$} \\
\midrule
\multirow{4}{*}{Latency}
 & Workload Prediction + Replica Sizing & 0.28/2.9 & 0.71/7.9 & 0.94/9.9 & \textbf{10.43}/4.2 & 0.99/99.0 \\
 & $+$ Forecast Calibration & 0.02/3.4 & 0.04/11.2 & 0.02/15.8 & 0.12/17.8 & 0.20/120.7 \\
 & $+$ Budget Pacing & 0.16/3.0 & 0.44/8.5 & 0.64/10.5 & 0.89/13.4 & 0.68/104.9 \\
 & Full BACC & 0.14/3.0 & 0.35/8.9 & 0.48/11.2 & 0.78/8.2 & 0.63/106.8 \\
\addlinespace
\multirow{4}{*}{CPU}
 & Workload Prediction + Replica Sizing & 0.70/3.1 & \textbf{3.65}/8.9 & \textbf{6.82}/11.0 & \textbf{14.05}/4.6 & \textbf{2.89}/111.7 \\
 & $+$ Forecast Calibration & 0.00/3.8 & 0.10/12.3 & 0.03/17.6 & 0.17/19.7 & 0.10/135.1 \\
 & $+$ Budget Pacing & 0.24/3.2 & 0.49/9.9 & 0.77/15.6 & 0.77/12.8 & 0.45/118.9 \\
 & Full BACC & 0.24/3.2 & 0.42/9.9 & 0.73/15.0 & 0.83/12.1 & 0.63/118.3 \\
\bottomrule
\end{tabular}
\end{table*}

Table~\ref{tab:unified_ablation} incrementally adds components to Replica Sizing. Prediction alone fails in 10/20 settings. Adding Forecast Calibration with a fixed risk tolerance, i.e., fixed over-provisioning, achieves compliance in all 20 but leaves 42--100\% of the budget unused. Budget Pacing reduces mean replicas by up to 35\% across all metric/target settings while preserving compliance, showing the benefit of adapting the risk tolerance to budget use.

Adding the Budget-Aware Trigger retains compliance, with mixed resource effects on A--C and E. On D, it reduces mean replicas in all four settings, including from 13.4 to 8.2 at the latency 99\% target.

\section{Kubernetes Validation}
\label{sec:generality}

We compare BACC and HPA in 24-hour Kubernetes replays at native trace rates, exercising live telemetry, controller decisions, Deployment updates, and pod readiness. Simulation uses the same trace and measurement windows.

\subsection{Testbed and Service}

We run a single-node kind cluster with Kubernetes v1.35.0 on an AMD EPYC 9754 server with 128 allocatable CPUs and 125~GiB of memory. The workload is a Go HTTP service whose request path performs a configurable amount of deterministic CPU work. Separate liveness and readiness endpoints keep health checking independent of request processing. Each application container has a 250-mCPU request and limit, a 64-MiB memory request, and a 256-MiB limit.

Every workload pod carries an Istio 1.30.3 sidecar. Destination-side Envoy histograms expose request counts and latency buckets to Prometheus, while Kubernetes reports application-container CPU and replica status. BACC runs in a separate Deployment and updates the workload's desired replica count. HPA uses the same service, load generator, and measurement path.

Controllable initialization work, readiness checks, and sidecar startup yield measured startup delays of approximately 62--68~s. Readiness is observed at second-level resolution; pending replicas do not serve requests. Exact software versions are listed with the released code.

\subsection{Trace Replay Protocol}

We replay the first evaluation day of regular HTTP trace~B and bursty, intermittent trace~D (Table~\ref{tab:trace_stats}). Open-loop requests use the original counts, spaced uniformly within each minute; missed sends are not carried forward. Request counts, client errors, and dispatch lag identify incomplete replays.

Both controllers start with the same replicas and use only available information. BACC uses five-minute regular decisions and one-minute observations. HPA reads mean application-container CPU at the default 15-second reconciliation cadence~\cite{kubernetes_hpa_docs}, with standard scale-up policy. Its 55\% CPU target and five-minute scale-down stabilization are selected on days~11--12 (Section~\ref{sec:cpu_eval}) and shared by both traces; no parameter is selected from live runs.

We repeat each trace--controller pair three times, for 12 complete replays. The first two repetitions use the same execution order, while the third reverses trace and controller order to expose systematic time-order effects.

An independent sampler queries the same destination-side Prometheus counters for both methods. Non-2xx responses and requests exceeding 500~ms violate the SLO. All reported metrics exclude a common five-minute warm-up, leaving 1,435 minutes. Resource use is mean desired replicas; mean Ready replicas are retained separately to quantify actuation effects.

\subsection{Results and Simulation Comparison}

\begin{table}[!t]
\centering
\caption{End-to-end Kubernetes replay and simulation over the same 24-hour
trace window after excluding a common 5-minute warm-up. Kubernetes entries
report mean $\pm$ sample standard deviation over three repetitions; Comp.
reports the number of compliant repetitions. $\bar R$ is mean provisioned
(desired) replicas. Bold
violation rates exceed the 5\% budget.}
\label{tab:k8s_validation}
\scriptsize
\setlength{\tabcolsep}{3pt}
\resizebox{\columnwidth}{!}{%
\begin{tabular}{llccc}
\toprule
Trace & Method & Simulation $S_{vr}$ / $\bar R$ & Kubernetes $S_{vr}$ / $\bar R$ & Comp. \\
\midrule
B & HPA  & 1.178\% / 7.269 & $(1.940 \pm 0.026)$\% / $7.899 \pm 0.027$ & 3/3 \\
B & BACC & 2.587\% / 6.372 & $(3.797 \pm 0.020)$\% / $6.580 \pm 0.011$ & 3/3 \\
\midrule
D & HPA  & \textbf{8.538\%} / 4.369 & \textbf{$(9.724 \pm 0.362)$\%} / $4.981 \pm 0.038$ & 0/3 \\
D & BACC & 4.352\% / 5.495 & $(4.337 \pm 0.785)$\% / $7.454 \pm 0.298$ & 3/3 \\
\bottomrule
\end{tabular}%
}
\end{table}

All 12 replays complete 1,440 minutes and attempt every scheduled request: 12,390,625 on B and 5,618,824 on D. No generator errors occur, and maximum dispatch lag is below 186~ms. Table~\ref{tab:k8s_validation} compares live means and sample standard deviations with the same-window simulation, not the 48-hour primary results. Unless noted, live results are repetition means. Variation is larger on D, but compliance outcomes agree across repetitions.

\paragraph{Regular trace B}
Both controllers satisfy the 5\% budget in 3/3 Kubernetes repetitions. HPA obtains the lower mean violation rate, 1.940\% versus BACC's 3.797\%. BACC uses 6.580 rather than 7.899 mean provisioned replicas; its mean paired replica reduction is 16.7\% across the three repetitions. The simulation predicts the same tradeoff: BACC uses 12.3\% fewer replicas while both methods comply. Desired and Ready replica means differ by 0.155 for BACC and 0.028 for HPA, consistent with modest actuation overhead on this regular workload.

\paragraph{Bursty trace D}
HPA complies in 0/3 repetitions, with mean violation rate 9.724\% and 4.981 provisioned replicas. BACC provisions more aggressively, averaging 7.454 replicas, and satisfies the budget in 3/3 repetitions at 4.337\%. Its mean paired bad-request reduction is 55.5\%, with a paired replica increase of 49.7\%.

The same-window simulation gives the same controller ordering: BACC reduces HPA's bad-request count by 49.0\%, while using 25.8\% more replicas. Its absolute rates are 4.352\% for BACC and 8.538\% for HPA. BACC therefore meets the budget in both environments, with margins of 0.648 and 0.663 percentage points in simulation and mean Kubernetes outcomes, respectively; HPA fails in both.

\paragraph{Sim-to-real behavior}
Live violation rates differ from simulation by $-0.015$ to $+1.210$ percentage points across the four cells; three are higher, while BACC on D is nearly identical. BACC's live replica mean is 3.3\% higher on B and 35.7\% higher on D, reflecting its response to variable readiness and harsher service outcomes.

On D, Figure~\ref{fig:k8s_trace_d_dynamics} shows HPA exceeding its cumulative budget early and remaining above pace. BACC enters the late burst at minute 1,380 below pace and retains a small final margin. The simulator uses fixed startup delay and a memoryless per-replica load surface; Kubernetes additionally exposes variable readiness, Prometheus window alignment, and recovery from overload and timeouts. Replay thus supports the comparative findings while identifying limits of the abstraction.

BACC completes 288 regular decisions per run. On D, it averages 24.3 early updates and changes desired replicas in 175.7 minutes, versus 205.0 HPA changes. On B, the corresponding counts are 89.3 early updates and 178.3 changes, versus 58.7 HPA changes. Trigger activity is thus workload-dependent.

\section{Conclusion}

BACC treats the violation allowance of a fixed-period SLO as a budget to spend over the period, not a limit to meet at every decision. It adjusts its risk tolerance to budget use: it over-provisions when little budget remains and accepts some violations when much remains. Its shared control loop supports both request-latency and CPU-threshold SLOs through metric-specific replica sizing. In simulation and Kubernetes replay, BACC meets the SLO with fewer replicas than fixed over-provisioning. Compliance holds across the tested forecasters, service profiles, random seeds, and startup-delay mismatch. These results support budget pacing as a practical resource-control mechanism within the evaluated settings.

\bibliographystyle{IEEEtran}
\bibliography{references}

\end{document}